\documentclass[prb,superscriptaddress,showpacs,preprintnumbers,amsmath,amssymb,floatfix,twocolumn]{revtex4}

\usepackage{graphicx}
\input{epsf}
\usepackage{dcolumn}
\usepackage{bm}
\begin{document}

\title{Thermodynamically self-consistent liquid state theories for systems with bounded potentials}

\author{Bianca M. Mladek}
\affiliation{Center for Computational Materials Science (CMS) and
  Institut f\"ur Theoretische Physik, TU Wien, Wiedner Hauptstra{\ss}e
  8-10, A-1040 Wien, Austria}
\affiliation{Institut f\"ur Experimentalphysik,
  Universit\"at Wien, Strudlhofgasse 4, A-1090 Wien, Austria}

\author{Gerhard Kahl}
\affiliation{Center for Computational Materials Science (CMS) and
  Institut f\"ur Theoretische Physik, TU Wien, Wiedner Hauptstra{\ss}e
  8-10, A-1040 Wien, Austria}

\author{Martin Neumann}
\affiliation{Institut f\"ur Experimentalphysik,
  Universit\"at Wien, Strudlhofgasse 4, A-1090 Wien, Austria}

\date{\today}

\begin{abstract}
  The mean spherical approximation (MSA) can be solved
  semi-analytically for the Gaussian core model (GCM) and yields --
  rather surprisingly -- exactly the same expressions for the
  energy and the virial equations. Taking advantage of this
  semi-analytical framework, we apply the concept of the
  self-consistent Ornstein-Zernike approximation (SCOZA) to the GCM: a
  state-dependent function $K$ is introduced in the MSA closure
  relation which is determined to enforce thermodynamic consistency
  between the compressibility route and either the virial or energy
  route.  Utilizing standard thermodynamic relations this leads to two
  different differential equations for the function $K$ that
  have to be solved numerically.  Generalizing our concept we propose
  an integro-differential-equation based formulation of the SCOZA
  which, although requiring a fully numerical solution, has the
  advantage that it is no longer restricted to the availability of an
  analytic solution for a particular system. Rather it can be used for
  an arbitrary potential and even in combination with other closure
  relations, such as a modification of the hypernetted chain
  approximation.
\end{abstract}

\maketitle

\section{Introduction}                 \label{introduction}

It is meanwhile well-known and widely documented that conventional
integral equation theories -- such as the Percus-Yevick (PY), the
hypernetted chain (HNC), or the mean spherical (MSA) approximation --
are thermodynamically inconsistent, which means that the various
thermodynamic routes to calculate the dimensionless equation of state
lead to significantly different results \cite{Han86,Cac96}.  Over the
past decades considerable effort has been devoted to the formulation
of thermodynamically self-consistent liquid state theories, which, in
turn, have led to an improved description of the structural and
thermodynamic properties of liquids with harshly repulsive
potentials. In the first generation of these concepts, such as the
Rogers-Young (RY) \cite{Rog84}, the modified hypernetted chain (MHNC)
\cite{Ros79}, or the Zerah-Hansen approach \cite{Han85}, simple
functions were introduced in the respective closure relations to the
Ornstein-Zernike (OZ) equation which use a pointwise adjustable but
not explicitly state-dependent parameter to interpolate between the
conventional closures. Since self-consistency was enforced for each
state point independent of the neighboring ones, we shall call this
approach {\it locally self-consistent}.  The concepts of the second
generation of the self-consistent liquid state theories were based on
more sophisticated ideas: the SCOZA scheme \cite{scoza} introduced an
explicitly state-{\it dependent} function in the MSA closure relation
to the OZ-equation in order to enforce thermodynamic self-consistency
between different thermodynamic routes; the hierarchical reference
theory (HRT) \cite{Par95}, on the other hand, successfully merged
ideas of microscopic liquid state theory and renormalization group
concepts.  In both these advanced liquid state approaches
thermodynamic consistency was enforced in the {\it entire} space of
system parameters, which we shall call {\it global self-consistency}.

In recent years, increasing effort has been devoted to investigations
of the structural and thermodynamic properties of soft matter systems
\cite{LikPR}.  The interactions in such systems either diverge weakly
or even remain finite (``bounded'') at short distances, i.e.,
when particles overlap. These potentials are commonly referred to as
soft potentials.  Initially, they were investigated by means of
conventional \cite{LikPR} and, more recently, by means of locally
self-consistent integral-equation theories \cite{Lan00,Fer00}.
However, the advanced liquid state theories mentioned above have not
been generalized to systems with soft potentials so far. The HRT
concept, for instance, relies on the known properties of a suitable
reference system.  While for systems with strongly repulsive
interactions the hard-core (HC) liquid represents an obvious and very
successful choice, no such reference system can be identified for
liquids with soft potentials. We therefore have to rule out HRT, at
least for the time being. 

On the other hand, applications of the SCOZA-concept
\cite{scoza} to liquid systems were up to now restricted to those
cases where the respective interactions can be expressed as a
combination of HC potentials with an adjacent linear combination of
Yukawa-tails (HCY-systems) \cite{Sch04}.  This restriction can be
traced back to the fact that the rather elaborate SCOZA-formalism
\cite{scoza} is intricately linked to the availability of the analytic
solution of the MSA for such a system \cite{Blu78}. From this point of
view the obvious counterpart of HCY systems in soft matter is the
Gaussian core model (GCM) \cite{Sti76}.  For this system the pair potential is
given by

\begin{equation}                                 \label{phi_gauss}
\Phi(r) = \varepsilon \exp[-(r/\sigma)^2],
\end{equation}
where $\varepsilon$ is an energy- and $\sigma$ a
length-parameter. 

Within the framework of the MSA (in the case of soft
potentials sometimes also termed random phase approximation), the
structural and thermodynamic properties of the GCM can to a large
extent be expressed semi-analytically \cite{Lan00,Lou00}.  In this
contribution we extend the SCOZA formalism to the GCM.

The GCM can be interpreted as a simple model to describe soft
matter. It has been pointed out that, for example, the effective
(coarse-grained) interaction between two isolated non-intersecting
polymer chains or dendritic macromolecules can, to a very good
approximation, be represented by the GCM potential
\cite{Lou00a,Goe04}. It is for two reasons that this potential
represents an ideal candidate to apply the concept of global
thermodynamic self-consistency to systems with soft potentials: {\bf
(i)} as we will show, two of the three traditional thermodynamic
routes, i.e., the energy and the virial route, coincide {\it exactly}
for the GCM within the MSA, a fact that, to the best of our knowledge,
has not been documented in literature so far; there is even evidence
\cite{Mla05b} that this also holds true for other systems with soft
potentials. Therefore, thermodynamic self-consistency has to be
enforced between two routes only, which considerably facilitates the
theory. {\bf (ii)} Using the analytic expressions given by the MSA to
enforce thermodynamic consistency for the GCM it is possible to
derive, via standard thermodynamic relations, either an ordinary (ODE)
or a partial (PDE) differential equation for the state-dependent
function introduced in the closure relation of the SCOZA
\cite{note1}. The ODE results from combining the virial and
compressibility route and can be solved for each isothermal line
independently. The PDE enforces consistency between the energy and
compressibility route and relates both density- and
temperature-derivatives. These two differential equations, although of
different complexity, have to be solved numerically and lead within
numerical accuracy to consistent results.
In addition, we propose an equivalent, integro-differential-equation
(IDE) based formulation of the SCOZA, which also has to be solved
numerically. This latter approach has the advantage that it can be
used for an arbitrary (soft) potential and in combination with closure
relations other than the MSA, such as e.g. a HNC-based SCOZA ansatz;
thus, it is no longer restricted to systems where semi-analytic
solutions of liquid state theories are known.

The rest of the paper is organized as follows: in Sec.~\ref{msa} we
re-visit the MSA for the GCM, providing thus the basis for the
(semi-)analytic formulation of the SCOZA.  In Sec.~\ref{scoza} we
present the ideas of SCOZA and derive the two differential equations
and the IDE that impose self-consistency and in Sec.~\ref{numerics} we
present details about the numerical solution strategies.
Sec.~\ref{results} is devoted to a detailed discussion of the SCOZA
results and a comparison with MC simulation data. Finally, in
Sec.~\ref{conclusion} we summarize and draw our conclusions.

\section{MSA}           \label{msa}

For the GCM, semi-analytic expressions for the
static and thermodynamic properties can be derived within the MSA \cite{Lan00,Lou00}.
Here we add a few details that have not been documented yet.

The MSA closure relation to the Ornstein-Zernike (OZ) equation,

\begin{equation}
h(r) = c(r) + \varrho \int {\rm d} {\bf r}' \, h(|{\bf r} - {\bf r}'|) \, 
c(r'), 
\end{equation}
where $h(r)$ and $c(r)$ are the total and the direct correlation
functions and $\varrho$ is the number density of the system, was
originally proposed for systems for which the pair potential consists
of a HC interaction with diameter $\sigma$ plus a tail that can
take different functional forms \cite{Han86}. For such potentials, the
MSA consists of an ansatz for $c(r)$,

\begin{equation}
c(r) = - \beta \Phi(r)  ~~~~~ {\rm for~}r>\sigma,
\end{equation}
where $\beta = (k_{\rm B}T)^{-1}$, $T$ is the temperature and $k_{\rm
B}$ Boltzmann's constant, along with the so-called core condition that
expresses the impenetrability of the particles

\begin{equation}               \label{corecondition}
g(r) = 0  ~~~~~ {\rm for~}r<\sigma.
\end{equation}
Here, $g(r) = h(r)+1$ is the radial distribution function (RDF).

As soft potentials lack a hard core, Eq.~(\ref{corecondition})
cannot be applied anymore and the MSA reduces to

\begin{equation}
c(r) = - \beta \Phi(r)  ~~~~~ {\rm for~all~}r.
\end{equation}

For the specific case of the GCM, where $\Phi(r)$ is a simple
Gaussian, this immediately leads to an analytic expression for the
static structure factor, $S(q)$,

\begin{equation}
S(q) =  \left[ 1 - \varrho \:\! \hat c(q) \right]^{-1}
= \frac{1}{1 + \alpha \exp[-(q^2 \sigma^2/4)]},
\end{equation}
where the hat denotes a Fourier transform, $q$ is the wave vector, and
$\alpha = \pi^{3/2} \varrho \sigma^3 \beta \varepsilon$. For the RDF
we find

\begin{equation}
g(r) = 1 - \frac{\alpha}{\varrho} \frac{1}{8 \pi^3}
\int d {\bf q} \,
{\rm e}^{-i {\bf q}{\bf r}} \frac{1}{{\rm e}^{q^2\sigma^2/4} + \alpha};
\end{equation}
in particular 

\begin{equation}                                  \label{g0}
g(0) = 1 + \frac{\beta \varepsilon}{\alpha} {\rm Li}_{3/2}(-\alpha).
\end{equation}
Here, ${\rm Li}_n(x)$ is the polylogarithm of order $n$ which is
discussed in detail in the Appendix.

Further, the thermodynamic properties of the GCM can be calculated
semi-analytically using one of the usual three thermodynamic routes
\cite{Han86}.  The results for the dimensionless equation of state,
$\beta P/\varrho$, where $P$ is the pressure, obtained via the
compressibility route ('C')

\begin{equation}
\left( \frac{\beta P}{\varrho} \right)^{\rm C} = 1 + \frac{1}{2}\:\! \alpha
\end{equation}
and the virial route ('V')

\begin{equation}                                  \label{p_v}
\left( \frac{\beta P}{\varrho} \right)^{\rm V} = 1 + \frac{1}{2}\:\! \alpha
-\beta \varepsilon \aleph(\alpha),
\end{equation}
where

\begin{equation}                                  \label{aleph}
\aleph(\alpha) = \frac{1}{2\alpha} \left[{\rm Li}_{3/2}(-\alpha) - 
{\rm Li}_{1/2}(-\alpha) \right]
\end{equation}
have already been reported in \cite{Lan00,Lou00}.

The energy route ('E') has not been considered in the literature so
far. To obtain $\left(\beta P/\varrho\right)^{\rm E}$ we first
calculate the excess (over ideal gas) internal energy per particle,
$U^{\rm ex}/N$,

\begin{equation}                                  \label{u_e}
\frac{\beta U^{\rm ex}}{N} = 2 \pi \beta \varrho \int\limits_0^\infty
{\rm d} r\, \Phi(r)\:\! g(r)\:\! r^2 = \frac{\alpha}{2} -
\frac{\beta \varepsilon}{2 \alpha} \left[ \alpha + {\rm
Li}_{3/2}(-\alpha) \right],
\end{equation}
from which we obtain the excess free energy per particle, $F^{\rm
ex}/N$,

\begin{equation}
\frac{\beta F^{\rm ex}}{N} = 
\int\limits_0^\beta \rm d\beta' \,\frac{U^{\rm ex}(\beta', \varrho)}{N} 
 = 
\frac{\alpha}{2} - \frac{\beta \varepsilon}{2 \alpha} 
\left[\alpha + {\rm Li}_{5/2}(-\alpha)\right],
\end{equation}
and, finally, the equation of state

\begin{equation}                                   \label{p_e}
\left( \frac{\beta P}{\varrho} \right)^{\rm E} = 1+ \varrho \:\!
\frac{\partial}{\partial \varrho} \left( \frac{\beta F^{\rm ex}}{N}
\right) = 1 + \frac{1}{2}\:\!\alpha -\beta \varepsilon \aleph(\alpha).
\end{equation}

Thus we find that virial (\ref{p_v}) and energy route (\ref{p_e}) lead
{\it exactly} to the same expressions for the dimensionless equation
of state. This is certainly an unexpected and atypical result. In
fact, in our numerical investigations of similar bounded systems in
combination with other closure relations, we have observed an
analogous, remarkable coincidence of the virial and the energy route
\cite{Mla05b}. To what extent this behavior is a general feature of
soft systems remains to be investigated. For those systems where
virial and energy route do coincide, this greatly facilitates the
formulation of thermodynamically self-consistent integral equation
theories, since consistency has to be enforced only either between the
virial and the compressibility or between the energy and the
compressibility route.

\section{SCOZA}                   \label{scoza}

The original formulation of the SCOZA for HC systems \cite{scoza} is
based on the MSA. It enforces the RDF to vanish inside the core and
for distances larger than the core diameter sets the direct
correlation function proportional to the potential; the
proportionality factor contains a state-dependent function that
imposes thermodynamic consistency.
Following the same scheme used to generalize the MSA to soft
potentials (cf.~Sec.~\ref{msa}), we modify the original SCOZA ansatz:

\begin{equation}                                  \label{clos_scoza_1}
c(r) = \beta \:\! K(\varrho, \beta) \:\! \Phi(r) ~~~~~ {\rm for~all~}r,
\end{equation}
where $K(\varrho, \beta)$ is an as yet undetermined, state-dependent
function.  As the MSA is recovered for $K(\varrho, \beta) \equiv -1$,
the present formulation of the SCOZA takes advantage of the
availability of the semi-analytic solution of the MSA for the GCM
presented in the preceding subsection.

Thus, closed expressions can be derived for the thermodynamic
properties within the SCOZA. To simplify the notation we introduce a
function $\tilde \alpha(\varrho, \beta) = \pi^{3/2} \varrho \sigma^3
\beta \varepsilon K(\varrho,\beta) = \alpha K(\varrho, \beta)$, which
is explicitly state-dependent, but for simplicity suppress the
arguments of $\tilde \alpha$ in the following.

According to the compressibility route the density derivative of the equation of state is given by 

\begin{equation}                                  \label{chi_comp}
\left(\chi^{\rm C}_{{\rm red}}\right)^{-1} = \left(\varrho k_\mathrm{B} T \chi_{T}^{\rm C}\right)^{-1} 
= 1-\varrho\:\! \hat{c}(0) = 1 - \tilde \alpha,
\end{equation}
where $\chi_{T}^{\rm C}$ is the isothermal compressibility and the
reduced isothermal compressibility $\chi^{\rm C}_{\rm red}$ is the
zero wave-vector value of the structure factor $S(q)$.

Further, following the virial route one finds the following expression
for the dimensionless equation of state
\begin{eqnarray}                                  \label{p_vir}
\left(\frac{\beta P}{\varrho}\right)^{\mathrm V} &=& 1 - \frac{2 \pi}{3}
\varrho \int \limits_0^\infty {\mathrm{d} r}\, r^3 \:\!
\frac{\mathrm{d} \beta \Phi(r)}{\mathrm{d} r}\:\! g(r) \nonumber\\
&=& 1 +
\frac{1}{2}\:\!\alpha - \frac{\beta \varepsilon}{2 \tilde
\alpha}\left[\mathrm{Li}_{5/2}(\tilde \alpha) -
\mathrm{Li}_{3/2}(\tilde \alpha)\right].
\end{eqnarray}
%
Finally, according to the energy route the dimensionless excess energy
per particle is given by

\begin{equation}
\frac{\beta U^{\rm ex}}{N} = \frac{1}{2}\:\!\alpha + \frac{\beta \varepsilon}{2 \tilde \alpha}
\left[{\mathrm{Li}}_{3/2} (\tilde \alpha)-\tilde \alpha \right].
\end{equation}
The energy and the virial route already coincide within the MSA and this also holds for the SCOZA. We
are therefore left with one single inconsistency, which can be removed
either via the virial/compressibility or via the
energy/compressibility route; both possibilities will be considered
in the following subsections.

\subsection{Virial- and compressibility-route}

We start by calculating the compressibility via the virial route,
$\chi_{T}^{\rm V}$, which is achieved by differentiating Eq.~(\ref{p_vir}) with respect
to $\varrho$,
\begin{widetext}
\begin{eqnarray}                                  \label{vir}
\left(\frac{\partial \beta P}{\partial \varrho}\right)^{\mathrm V} 
& = & 1 + \alpha - \frac{1}{2 \sigma^3 \pi^{3/2} K(\varrho, \beta)^2}
\left\{ \frac{K(\varrho, \beta)}{\varrho} \left[\mathrm{Li}_{3/2}(\tilde
\alpha) - \mathrm{Li}_{1/2}(\tilde \alpha)\right]\right. \nonumber\\ 
 & &
+\left.\frac{\partial K(\varrho, \beta)}{\partial \varrho}
\left[2\mathrm{Li}_{3/2}(\tilde \alpha)-\mathrm{Li}_{5/2}(\tilde \alpha)-
\mathrm{Li}_{1/2}(\tilde \alpha)\right]\right\}.
\end{eqnarray}
\end{widetext}
Equating this result with the compressibility as obtained via the
compressibility route (\ref{chi_comp}) leads to the following ODE for
the unknown function $K(\varrho, \beta)$:

\begin{widetext}
\begin{equation}                                  \label{scoza_pde_1}
\frac{\partial K(\varrho,\beta)}{\partial \varrho} = \frac{K(\varrho, \beta) 
\left\{2 \pi^3 \beta \varepsilon \varrho^2 \sigma^6 K(\varrho, \beta)\,
\left[K(\varrho, \beta)+1\right] - 
\left[{\mathrm{Li}}_{3/2}(\tilde \alpha) - 
{\mathrm{Li}}_{1/2}(\tilde \alpha)\right]\right\}}
{\varrho\left[2{\mathrm{Li}}_{3/2}(\tilde \alpha)-
{\mathrm{Li}}_{5/2}(\tilde \alpha)-
{\mathrm{Li}}_{1/2}(\tilde \alpha)\right]}.
\end{equation}
\end{widetext}
Note that this ODE can be solved for each isothermal line separately.

Analyzing the ODE, we note that the right hand side (RHS) of
Eq.~(\ref{scoza_pde_1}) contains two singularities. Obviously the
denominator vanishes for $\varrho \to 0$, but expanding numerator and
denominator around $\varrho=0$, we find that

\begin{equation}                                \label{inital_cond}
K(\varrho=0;\beta) = - \frac{4 \sqrt{2}}{4 \sqrt{2} + \beta \varepsilon}.
\end{equation}
Further, the denominator also vanishes at $\tilde \alpha = \tilde
\alpha_0 \approx -7.7982$. This, however, turns out to be a removable
singularity which can be treated by appropriate means (cf.~subsection
\ref{sec_ode}).

\subsection{Energy- and compressibility-route}

To enforce thermodynamic consistency between the energy and
compressibility route we utilize the variant of the SCOZA-formalism
proposed in \cite{Pin98,Sch02} which brought along a breakthrough of
this concept for systems with repulsive potentials. This approach is
based on replacing the differential equation for $K(\varrho, \beta)$
by one for the excess energy density $u=U^{\rm ex}/V$. To this end, we
consider the following thermodynamic relation (see, e.g.,
\cite{Cac99})

\begin{equation}                                  \label{rel_u_chi}
\frac{\partial}{\partial \beta}
\left(\frac{1}{\chi_{\mathrm{red}}^{\rm E}}\right) = \varrho\:\!
\frac{\partial^2 u}{\partial \varrho^2}.
\end{equation}
Expressing at constant density $\chi_{\rm red}^{\rm E}$ as a function
of $u$, the left hand side can be rewritten as

\begin{equation}
\frac{\partial}{\partial \beta}
\left(\frac{1}{\chi_{\mathrm{red}}^{\rm E}(u)}\right) =
\frac{\partial}{\partial u} \left(\frac{1}{\chi_{\mathrm{red}}^{\rm
E}(u)}\right) \frac{\partial u}{\partial \beta},
\end{equation}
so that finally Eq.~(\ref{rel_u_chi}) becomes

\begin{equation}                \label{scoza_pde_2}                      
\frac{\partial u}{\partial \beta} = \left[\frac{\partial}{\partial u}
\left(\frac{1}{\chi_{\mathrm{red}}^{\rm E}(u)}\right)\right]^{-1} 
\varrho\:\! \frac{\partial^2 u}{\partial \varrho^2}.
\end{equation}
In contrast to Eq.~(\ref{scoza_pde_1}), this relation contains
derivatives with respect to both $\varrho$ and $\beta$ and is a PDE of
the diffusion type. However, the diffusivity, $D(\varrho, \beta) =
\left[\frac{\partial}{\partial u} \varrho
\left(\frac{1}{\chi_{\mathrm{red}}(u)}\right)\right]^{-1}$, is
state-dependent \cite{note2} and turns out to be negative which
renders the numerical solution extremely
intricate. $\chi_{\mathrm{red}}(u)$ is now identified with the
expression obtained by the compressibility route (\ref{chi_comp})

\begin{equation}
\left[\chi_{\mathrm{red}}^{\rm C}(u)\right]^{-1} = 1 -
\tilde \alpha,
\end{equation}
where $K(u)$ is determined by inverting the result of the energy route

\begin{equation}                             \label{inversion}
u = \frac{\varrho}{\beta} \left\{ \frac{1}{2}\:\!\alpha+
\frac{\beta \varepsilon}{2 \tilde
\alpha}\left[\mathrm{Li}_{3/2}(\tilde \alpha) - \tilde
\alpha\right]\right\}.
\end{equation}

\subsection{Integro-differential equation approach}    \label{scoza_e}
 
So far, in deducing the SCOZA-ODE (\ref{scoza_pde_1}) and PDE
(\ref{scoza_pde_2}), we have taken advantage of the availability of
the semi-analytic framework provided by the MSA for the properties of
the GCM. Unfortunately, this represents a rather singular
exception. In order to eliminate the restrictions resulting from this
fact one may ask whether the SCOZA-concept may be formulated for the
{\it general} case, i.e., when a semi-analytic solution to the
MSA is not at hand. This is indeed possible as we show in the
following:
let us assume a SCOZA-type closure relation, i.e.,

\begin{equation}                                  \label{clos_scoza_2}
c(r) = \beta\:\! \bar K\:\! \Phi(r) ~~~~~ {\rm for~all~}r.
\end{equation}
Once $\bar K$ is specified, this leads in combination with the OZ
equation directly to the radial distribution function $g(r) = g(r; \varrho,
\beta; \bar K)$, which is thus also a function of $\bar K$. If we
assume $\bar K$ to be explicitly state-dependent, i.e., $\bar K = \bar
K(\varrho, \beta)$, the compressibility as determined by the virial
route, i.e., differentiating the standard virial equation of
state, is

\begin{widetext}
\begin{eqnarray}                                  \label{p_vir_2}
\left[\varrho k_{\rm B}T \chi_{T}^{\rm V} \right]^{-1} & = & 1 -
\frac{4\pi}{3} \varrho \int\limits_0^\infty {\rm d}r\, r^3\:\!
\frac{\mathrm{d} \beta \Phi(r)}{\mathrm{d}r}\:\!  g(r; \varrho, \beta;
\bar K) - \frac{2\pi}{3} \varrho^2 \int\limits_0^\infty {\rm d}r\,
r^3\:\!  \frac{\mathrm{d} \beta \Phi(r)}{\mathrm{d}r}\:\!
\frac{\partial g(r; \varrho, \beta; \bar K)}{\partial \varrho}
\nonumber \\ & & - \frac{2\pi}{3} \varrho^2 \frac{\partial \bar
K}{\partial \varrho} \int\limits_0^\infty {\rm d}r\, r^3\:\!
\frac{\mathrm{d} \beta \Phi(r)}{\mathrm{d}r}\:\!  \frac{\partial g(r;
\varrho, \beta; \bar K)}{\partial \bar K}.
\end{eqnarray}
\end{widetext}
Thermodynamic self-consistency between the virial and the
compressibility route is now enforced by choosing the mixing
parameter $\bar K$ such that $\chi^{\rm C}_{T}$ is equal to $\chi^{\rm
V}_{T}$, i.e., by finding at fixed temperature $T$ a root of the
function

\begin{equation}                                  \label{consist_f}
f(\bar K) = \chi^{\rm C}_{T} - \chi^{\rm V}_{T}.
\end{equation}
Here, derivatives with respect to $\varrho$ and $\bar K$ have to be
calculated numerically (see below).

It was exactly this idea that was realized in previous applications of
parameterized closure relations such as RY \cite{Rog84}, HMSA
\cite{Han85}, or MHNC \cite{Ros79}. There, however, consistency was
then achieved only {\it locally}, i.e., considering each state point
in isolation and neglecting thus the density dependence of $\bar K$.
This corresponds to setting $\partial \bar K / \partial \varrho=0$ and
dropping the last term in Eq.~(\ref{p_vir_2}). In the present approach, in
contrast, since we consider $\bar K$ to be explicitly state-dependent,
this term is retained. Thus, the consistency criterion involves not
only isolated state points but also via the density derivative nearby
state points; therefore we have chosen to call the criterion a {\it global}
one. The quantitative difference between the local and the global
approaches will be discussed below.

\subsection{HNC-based SCOZA}

The above approach to thermodynamic consistency represents an IDE
based re-formulation of the SCOZA; it is not only entirely independent of the
semi-analytic solution provided by the MSA for the GCM and thus becomes completely
{\it general} in the sense that in this formulation
self-consistency can now be enforced for systems with {\it arbitrary}
(soft) potentials and in combination with other, parameterized closure
relations.

To demonstrate the power of this idea we introduce a HNC-based
SCOZA (for clarity we will refer to the SCOZA approach introduced
above ``conventional SCOZA''). This particular choice is motivated by
the fact that the HNC has been found to work very well for the GCM and
other soft potentials \cite{Lan00,Lou00}. For the closure relation of
our HNC-based SCOZA we propose

\begin{equation}                                     \label{hnc_scoza}
g(r) = \exp(\beta \bar K_{\mathrm{HNC}}(\varrho, \beta) \Phi(r) + h(r) - c(r)),
\end{equation}
where the unknown, state-dependent function $\bar
K_{\mathrm{HNC}}(\varrho, \beta)$ is determined such as to make the
RHS of the consistency requirement (\ref{consist_f}), i.e., the
equality between the compressibility and the virial route,
vanish. Numerically, this is achieved by solving Eq.~(\ref{p_vir_2}),
where $g(r; \varrho, \beta; \bar K_{\mathrm{HNC}})$ is now obtained from the
solution of the OZ-equation along with the closure relation
(\ref{hnc_scoza}).

\section{Numerical solution of the SCOZA and computer simulations}    \label{numerics}

\subsection{ODE approach}                          \label{sec_ode}

The SCOZA-ODE (\ref{scoza_pde_1}) has the attractive feature that it
can be solved for each isothermal line independently and represents the
fastest route among the three alternative formulations presented above
to determine $K(\varrho, \beta)$. We have used an implicit
fourth-order Runge-Kutta algorithm \cite{NR92}
to solve this ODE numerically. This works generally very well except
for those state points where the expression in the square brackets of
the denominator of the RHS of Eq.~(\ref{scoza_pde_1}) vanishes. A closer
analysis shows that this singularity at $\tilde \alpha = \tilde
\alpha_0$ is removable, since both numerator and denominator vanish
simultaneously. In fact, splitting the density range in two regions,
depending on whether $\tilde \alpha$ is smaller or larger than $\tilde
\alpha_0$, and integrating the ODE ``forward'' [starting at $\varrho =
0$ with initial value (\ref{inital_cond})] in the former and
``backward'' (from a sufficiently high density so that $K=-1$) in the
latter, we were able to smoothly join the partial solutions at $\tilde
\alpha = \tilde \alpha_0$ and thus to obtain $K(\varrho,\beta)$ over
the entire density range.
We point out that reliable solutions of this ODE can only be obtained
if an efficient and accurate evaluation of the polylogarithm is
guaranteed (see Appendix). $K(\varrho, \beta)$ as a function of
$\varrho$ and $\beta$ is displayed in Figure \ref{fig_k}; the results
will be discussed in Sec.~\ref{results}.

Alternatively, we have also solved this ODE with {\tt MATHEMATICA}
using a Livermore solver for ordinary differential equations with
automatic method switching (LSODA) \cite{Wol03}. The polylogarithms
encountered in the RHS of Eq.~(\ref{scoza_pde_1}) are evaluated in {\tt
MATHEMATICA} with high accuracy (for details see
\cite{Wol03}). Altough the differential-equation-solver package is not
able to deal properly with the removable singularity noted above and
breaks down for $\tilde \alpha \sim \tilde \alpha_0$, outside this
small range, {\tt MATHEMATICA} provides quasi-exact reference data for
the function $K(\varrho, \beta)$.

\subsection{PDE approach}

From the numerical point of view, solving the diffusion-type SCOZA-PDE
(\ref{scoza_pde_2}) is a delicate task. Boundary conditions at
$\varrho =0$ and at a large, but finite $\varrho_{\rm max}$, as well
as an initial condition at $\beta = 0$ are required. In particular the
boundary condition at $\varrho_{\rm max}$ has to be chosen
carefully. In contrast to HC systems, where boundary conditions follow
naturally from the existence of a maximum density, particles that
interact via bounded potentials can fully overlap and thus it is
possible to compress the system to arbitrary high densities. For this
region, we know that the MSA becomes exact and thus self-consistent
\cite{Lan00}, i.e., $K(\varrho=\infty, \beta) = -1$.  In numerical
calculations, however, we are forced to set $K = -1$ at some finite
maximum density $\varrho_{\rm max}$. Furthermore, we have to face the
problem of a state-dependent diffusivity $D(\varrho,\beta)$. Since
this quantitiy is even negative, not only the solution to the PDE but
any numerical error incurred in obtaining it may be expected to grow
exponentially. Among other things, small errors made in the
formulation of the boundary conditions and the inversion of the highly
non-linear relation (\ref{inversion}) to determine $D(\varrho, \beta)$
will eventually get dominant. Together, these difficulties make it
practically impossible to reliably solve this PDE. Taking on the other
hand $K(\varrho, \beta)$ as obtained from the ODE (\ref{scoza_pde_1})
and inserting it into Eq.~(\ref{scoza_pde_2}), we find that this
relation is fulfilled very accurately, which proves the numerical
consistency of the two differential equations approaches to the SCOZA.

\subsection{IDE approach}

The IDE based formulation of the SCOZA, i.e., equations
(\ref{consist_f}) along with (\ref{p_vir_2}), has been solved
iteratively using both the conventional (\ref{clos_scoza_2}) and the
HNC-based closure (\ref{hnc_scoza}). We introduce a density-grid (with
spacing $\Delta \varrho$) and assume a starting value $\bar K =
-1$. We solve the OZ equation with the appropriate closure relation
using standard integral-equation solver algorithms for a given state
point (i.e., we fix $\varrho$ and $\beta$) and the neighboring
density-values (i.e., for $\varrho \pm \Delta \varrho$). Thus, the
derivatives in the RHS of Eq.~(\ref{p_vir_2}) can be calculated
numerically. Due to the appearance of the derivative $\partial \bar
K/\partial \varrho$, Eq.~(\ref{p_vir_2}) has to be solved iteratively
and leads then to $\bar K(\varrho, \beta)$ for the {\it entire}
density range considered. As a consequence of the iterative and purely
numerical character of the solution strategy, this approach is more
time consuming than the solution of the ODE (\ref{scoza_pde_1}).

\subsection{Monte Carlo simulations}

To test the reliability of our integral equation results we have
generated reference data for the GCM by means of standard Monte Carlo
(MC) simulations in the canonical ensemble. For each thermodynamic
state considered, we started from a random configuration of ${\cal N}
= 1\:\!  000$ particles. The system was at first allowed to
equilibrate for $10\,000$ passes, where a pass consists of ${\cal N}$
trial moves, i.e., on average each particle has been subjected to a
trial move once. After that, we have carried out production runs of
another $150-300\,000$ passes to calculate the desired ensemble
averages.

\section{Results}                        \label{results}

\begin{figure}
\begin{center}
\includegraphics[width=7.cm,clip]{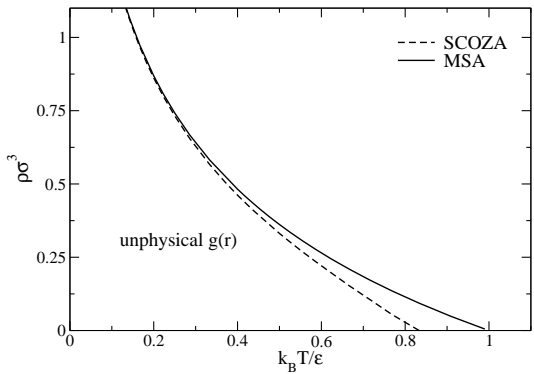}
\end{center}
\caption{Region in the density-temperature-space, where the MSA and
the SCOZA provide unphysical results, i.e., the RDF $g(r)$ attains
negative values.}
\label{fig_limits}
\end{figure}

We start the discussion of our results by specifying the range in
$(\varrho, \beta)$-space where the MSA and the SCOZA provide unphysical
results, i.e., where $g(r)$ is negative (see Figure
\ref{fig_limits}). While this failure of the MSA was briefly addressed
in \cite{Lou00}, we think that a more {\it quantitative} analysis is
in order, since similar problems might be encountered in applications
of the MSA (and of related concepts) to other systems with soft
potentials. In fact, also for the SCOZA unphysical results can be
obtained for certain system parameter combinations. For the MSA the
limits of this range of unphysical behaviour are easily determined via
Eq.~(\ref{g0}), and for the SCOZA they are found from the
equivalent, generalized expression (i.e., replacing $\alpha$ by
$\tilde \alpha$). Results are shown in Figure \ref{fig_limits},
indicating that at low temperatures the MSA and the SCOZA both become
unphysical if the density is reduced below some threshold density
$\varrho = \varrho(\beta)$. It is interesting to note that similar
problems of unphysical solutions and thus restricted applicability
have also been reported for other self-consistent schemes, such as the
RY- or the zero-separation concepts, in combination with the GCM
\cite{Lan00}.

\begin{figure}
\begin{center}
\includegraphics[width=7.cm,clip]{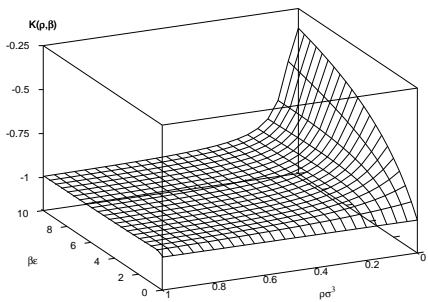}
\end{center}
\caption{ $K(\varrho,\beta)$ as obtained from the solution of the
  SCOZA-PDE (\ref{scoza_pde_1}) over a representative range of
  $\varrho \sigma^3$ and $\beta \varepsilon$. Note that $\beta
  \varepsilon$ is the inverse reduced temperature, i.e., high values
  of $\beta \varepsilon$ correspond to low temperatures.}
\label{fig_k}
\end{figure}

The state-dependent function $K(\varrho, \beta)$ which guarantees in
the SCOZA-scheme {\it full} thermodynamic self-consistency, i.e.,
between all three thermodynamic routes is depicted in Figure
\ref{fig_k} in a representative part of the parameter space. Detailed numerical investigations have shown that all three SCOZA
formulations presented in the previous chapters provide -- within
numerical accuracy and despite different levels of numerical
complexity -- equivalent results.

Bearing in mind that the MSA is recovered for $K(\varrho, \beta)
\equiv -1$, we observe that this function differs substantially from
this value at low densities (with a pronounced
temperature-dependence), thus indicating those regions where the MSA
is thermodynamically inconsistent. At high densities we confirm
earlier results reported in \cite{Lan00,Lou00}, which have stated that
in this regime the MSA becomes exact and thus self-consistent.  While
in \cite{Lou00} this conclusion was based on an analysis of the large
density-behaviour of the function $\aleph$ as defined in
Eq.~(\ref{aleph}), our argumentation follows directly from a visual
inspection of the function $K(\varrho, \beta)$.

\begin{figure}
\begin{center}
\includegraphics[width=7.cm,clip]{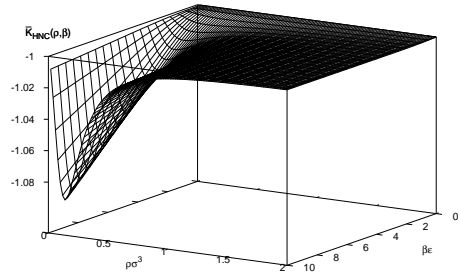}
\end{center}
\caption{ $\bar K_{\mathrm{HNC}}(\varrho,\beta)$ as obtained from the
  solution of the IDE approach to the HNC-based SCOZA closure
  (\ref{hnc_scoza}) over a representative range of $\varrho \sigma^3$
  and $\beta \varepsilon$. Note that, in an effort to enhance the
  visibility of the data, the viewpoint is now different from the one
  in Figure \ref{fig_k}.}
\label{fig_hnc_scoza_1}
\end{figure}

For the HNC-based SCOZA, the corresponding function, $\bar
K_{\mathrm{HNC}}(\varrho, \beta)$ is shown in Figure
\ref{fig_hnc_scoza_1}. Taking the deviation of this function from $-1$
as a measure of the thermodynamic inconsistency of the simple
HNC-approach (similar to the case of the MSA), we observe that the HNC is
to a large degree self-consistent. It is only at small densities and
low temperatures that $\bar K_{\mathrm{HNC}}(\varrho, \beta)$ slightly
deviates from $-1$. This large degree of thermodynamic
self-consistency of the HNC for systems with bounded potentials was
already observed for selected state points in \cite{Lou00, Lan00}, but
was never demonstrated on a quantitative level for a wider range of
system parameters.

\begin{figure}
\begin{center}
\includegraphics[width=7.cm,clip]{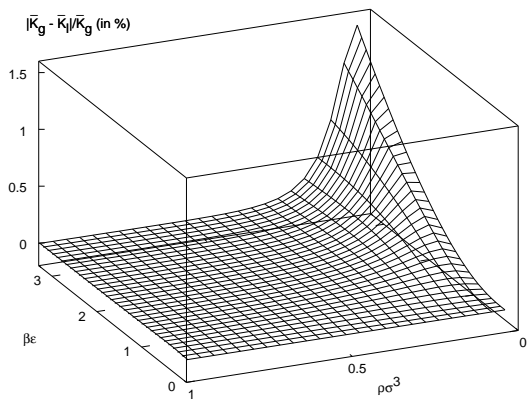}
\end{center}
\caption{ Relative difference between the functions $\bar K_g$ and
  $\bar K_l$ (as defined in the text) over a representative range of
  $\varrho \sigma^3$ and $\beta \varepsilon$.}
\label{fig_k_loc_glob}
\end{figure}

\begin{figure}
\begin{center}
\includegraphics[width=7.cm,clip]{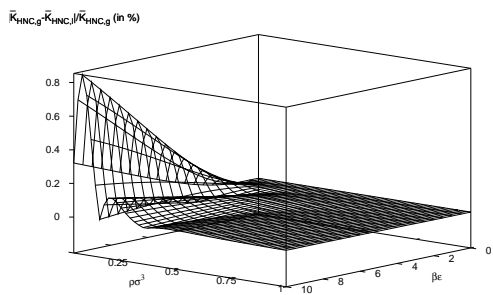}
\end{center}
\caption{ Relative difference between the functions $\bar
  K_{\mathrm{HNC},g}$ and $\bar K_{\mathrm{HNC},l}$ (as defined in the
  text) over a representative range of $\beta \varepsilon$ and
  $\varrho \sigma^3$.  Note that, in an effort to enhance the
  visibility of the data, the viewpoint is now different from the one
  in Figure \ref{fig_k_loc_glob}.}
\label{fig_k_loc_glob_hnc}
\end{figure}

We conclude our discussion of the thermodynamic self-consistency of
the conventional (MSA-based) and the HNC-based SCOZA concepts by a
direct comparison between local and global self-consistency, as
defined in subsection \ref{scoza_e}. Let $\bar K_g(\varrho,\beta)$
denote the explicitly state-dependent function $\bar K(\varrho,
\beta)$, as introduced to enforce thermodynamic self-consistency in
the IDE formulation of the (MSA- or HNC-based) SCOZA
(cf.~Sec.~\ref{scoza_e}); thus, the subscript '$g$' stands for {\it
global} self-consistency. On the other hand, if the last term in
Eq.~(\ref{p_vir_2}) is neglected thermodynamic self-consistency is
only enforced for a single, isolated state point and in this case we
denote the function by $\bar K_l(\varrho, \beta)$ ({\it local}
self-consistency). In Figure \ref{fig_k_loc_glob} we show the relative
difference between these functions for the conventional SCOZA and we
observe that it amounts to a few percent only for small densities,
even down to intermediate temperatures. Figure
\ref{fig_k_loc_glob_hnc} shows the same function for the HNC-based
SCOZA. Here, the differences become noticeable only at small densities
and low temperatures. Thus, over a large parameter range {\it local}
consistency is in both cases already a good substitute for {\it
global} consistency.

\begin{figure}
\begin{center}
\includegraphics[width=7.cm,clip]{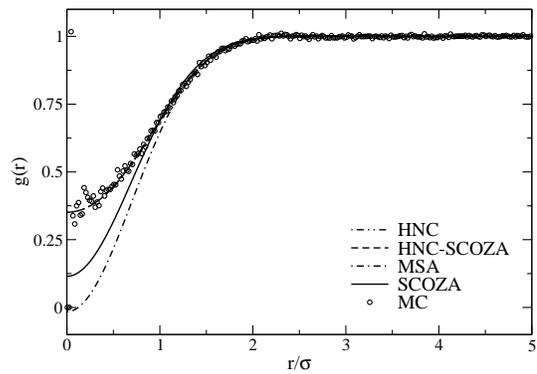}
\end{center}
\caption{RDF $g(r)$ for the GCM for $\beta \varepsilon = 1.1$ and
  $\varrho \sigma^3 = 0.04$. Note that the HNC and the HNC-based SCOZA
  curves coincide within line thickness.}
\label{fig_pdfs}
\end{figure}

\begin{figure}
\begin{center}
\includegraphics[width=7.cm,clip]{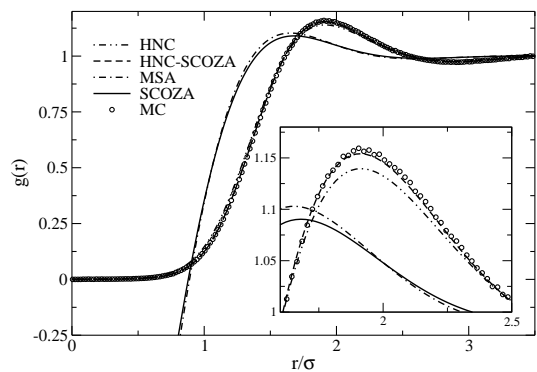}
\end{center}
\caption{RDF $g(r)$ for the GCM for $\beta \varepsilon = 10$ and
  $\varrho \sigma^3 = 0.14$.}
\label{fig_pdfs_a}
\end{figure}

We now turn to the structural properties of the GCM by comparing the
RDFs for two different thermodynamic states. In Figure \ref{fig_pdfs},
we have chosen a state-point close to the boundary where the SCOZA
becomes unphysical (cf.~Figure \ref{fig_limits}). We observe that
compared to the MC reference data, the conventional SCOZA does bring
along a slight improvement over the MSA. On the other hand, the
results provided by the HNC and the HNC-based SCOZA both reproduce the
MC-data perfectly. Figure \ref{fig_pdfs_a} shows the RDF for the GCM
at a low temperature and low density. Here, we are in the regime where
both the MSA and the MSA-based SCOZA provide unphysical results. We
see that while the conventional HNC results already reproduce the MC
data rather well, the HNC-based SCOZA leads to a perfect agreement
with the simulations. We conclude, that although the MSA-based SCOZA
for the GCM does not bring along the same improvement for the
structural properties as documented for HCY-systems, the concept of
self-consistency by itself proves to be of great value when used
with a closure better adapted to bounded potentials, i.e., a HNC-based
closure.

\begin{figure}
\begin{center}
\includegraphics[width=7.cm,clip]{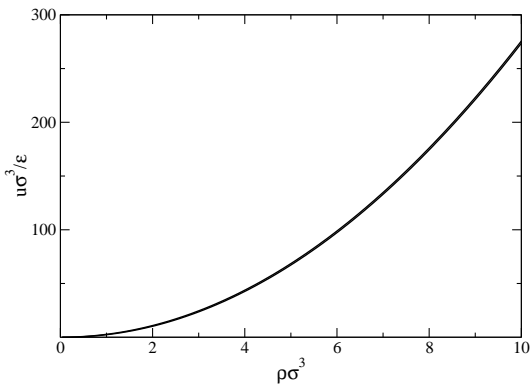}
\end{center}
\caption{$u \sigma^3/\varepsilon$ calculated within the conventional
  SCOZA as a function of $\varrho \sigma^3$ for different inverse
  reduced temperatures $\beta \varepsilon \in [0, 10]$; for discussion
  see text.}
\label{fig_u_univ}
\end{figure}

Finally, we examine some thermodynamic properties and start our
discussion by presenting a rather surprising result: if we plot the
quantity $u \sigma^3/\varepsilon$ as a function of the density, then
the curves evaluated for different isothermal lines practically
coincide (cf.  Figure \ref{fig_u_univ}); even though we present only
the conventional SCOZA, we note that this coincidence is also observed
for the MSA, the HNC, and the HNC-based SCOZA. This remarkable scaling
behavior might be worth being the subject of future investigations.

\begin{figure}
\begin{center}
\includegraphics[width=7.cm,clip]{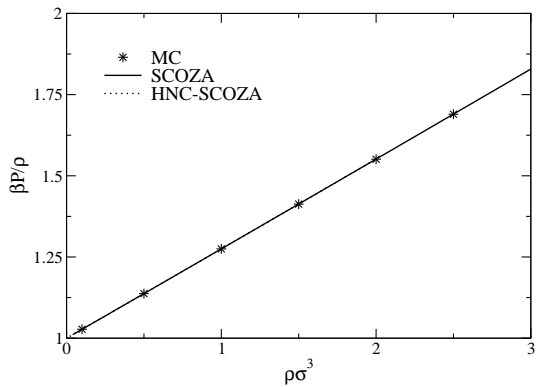}
\end{center}
\caption{$\beta P/\varrho$ as a function of $\varrho \sigma^3$ for the
  GCM for $k_{\rm B} T/\varepsilon=10$. The results of the SCOZA and
  the HNC-based SCOZA coincide. Both SCOZA approaches provide physical
  data for the RDF (i.e., $g(r)>0$) over the entire density-range.}
\label{fig_p_1}
\end{figure}

\begin{figure}
\begin{center}
\includegraphics[width=7.cm,clip]{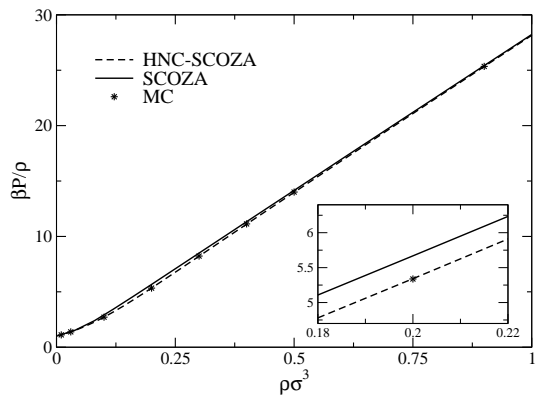}
\end{center}
\caption{ $\beta P/\varrho$ as a function of $\varrho \sigma^3$ for the
GCM for $k_{\rm B} T/\varepsilon=0.1$. Note that the conventional
SCOZA provides unphysical results for the RDF (i.e., $g(0)< 0$) for $\varrho \sigma^3
\lesssim 1.27$.}
\label{fig_p_2a}
\end{figure}

\begin{figure}
\begin{center}
\includegraphics[width=7.cm,clip]{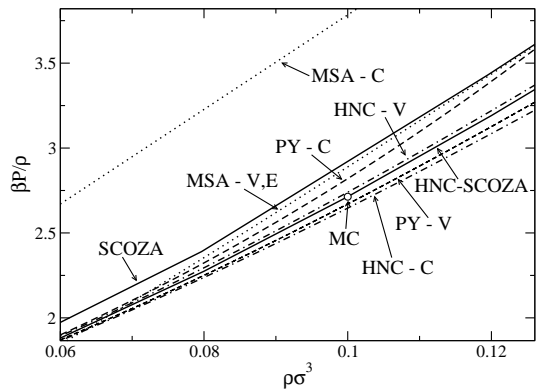}
\end{center}
\caption{Same as Figure \ref{fig_p_2a}, showing an enlarged view of a
limited $\varrho \sigma^3$-range. Lines and symbols as labeled.}
\label{fig_p_2b}
\end{figure}

We conclude this section with the results for the dimensionless
equation of state, $\beta P/\varrho$, for two different temperatures,
i.e., $k_{\rm B}T/\varepsilon = 10$ (see Figure \ref{fig_p_1}) and
$k_{\rm B}T/\varepsilon = 0.1$ (see Figures \ref{fig_p_2a} and
\ref{fig_p_2b}).  For $k_{\rm B}T/\varepsilon = 10$, we find that the
SCOZA-results coincide with high accuracy with the MC-data. For
$k_{\rm B}T/\varepsilon = 0.1$ we observe (Figure \ref{fig_p_2a}) that
the conventional SCOZA provides data that are obviously very close to
those obtained by simulations, while the HNC-based SCOZA data fit them
perfectly. A more thorough comparison, including this time also other
liquid state theories, such as the MSA, the PY, or the HNC
approximations \cite{Han86}, is displayed on an enlarged scale in
Figure \ref{fig_p_2b}. We observe that in addition also the virial
route of the PY and of the HNC (as expected \cite{Lan00,Lou00})
nicely reproduce the MC data; however, while the SCOZA is
self-consistent, this is not the case for the conventional closure
relations HNC and PY: their respective compressibility-data sometimes
differ distinctively from their virial and/or energy results.

Thus we can conclude that the general concept of the SCOZA {\it does}
bring along an improvement over conventional liquid state theories for
a thermodynamically consistent description of the properties of the
GCM, in particular if used in combination with a modification of the
HNC closure. It is especially remarkable that also the structural
properties are enhanced, even though the SCOZA scheme only enforces
self-consistency for the thermodynamic properties.

\section{Conclusion}                 \label{conclusion}

Motivated by the success of the SCOZA to describe the properties of HC
systems, we have made first steps to extend this concept to systems
with soft potentials. The fact that the MSA can be solved
semi-analytically for the GCM makes this system an ideal candidate for
a first application of the SCOZA.
 
Due to the fact that virial and energy route happen to yield exactly
the same result for the GCM within the MSA (and possibly also for
other closure relations), we are left to fix only one inconsistency,
namely between one of these two routes and the compressibility route
on the other hand side. Introducing a state dependent function
$K(\varrho, \beta)$ in the MSA closure, we were able to derive three
different approaches, namely an ODE, a PDE, and an IDE, that enforce
thermodynamic self-consistency. While the ODE and PDE rely on the
analytic solution provided by the MSA for this particular system, the
IDE formulation is completely independent of this framework and can be
applied for arbitrary systems and in combination with any closure
relation. It remains to be verified whether the IDE approach is also
applicable to systems with repulsive potentials.

The three formulations provide results for $K(\varrho, \beta)$ and
$\bar K(\varrho, \beta)$ that are equivalent within numerical
accuracy. In contrast to systems with harshly repulsive potentials the
improvement of the conventional SCOZA approach over the MSA data is
less spectacular. While it coincides with the MSA results in the
limiting case of high densities where the MSA is already
self-consistent, the (conventional) SCOZA represents a substantial
improvement at small densities and low temperatures where the
thermodynamic inconsistency of the MSA is more pronounced. Replacing
the conventional SCOZA relation by a HNC-type closure that contains an
analogous state-dependent function $\bar K_{\mathrm{HNC}}(\varrho,
\beta)$, we are able to improve the HNC-data for the structural as
well as for the thermodynamic properties of the system. With this
generalised approach we have not only demonstrated the flexibility and
power of the IDE approach but have also proposed what may turn out to
become a reliable liquid state theory for systems with bounded
potentials.

\section{Appendix}

The polylogarithm of order $n$, ${\rm Li}_n(z)$, also known as
Jonqui{\`e}re's function \cite{Mmaxx}, is a complex valued
function of complex argument $z$, defined by

\begin{equation}                                  \label{poly_n}
{\rm Li}_n(z) = \frac{z}{\Gamma(n)} \int\limits_0^\infty {\rm d}t \,
\frac{t^{n-1}}{{\rm e}^t-z},
\end{equation}
where $n$ is a positive, real parameter. If $z \in \mathbb{R}
\backslash (1, \infty)$, then the polylogarithm is real-valued
\cite{42DA}. For $|z|<1$ the polylogarithm can be evaluated as a power
series

\begin{equation}                                 \label{poly_sum}
{\rm Li}_n(z) = \sum_{k=1}^\infty \frac{z^k}{k^n}.
\end{equation} 
A relation that turned out to be useful for the present application is

\begin{equation}
\frac{\mathrm{d}}{\mathrm{d}z} {\rm Li}_n(z) = \frac{1}{z} {\rm Li}_{n-1}(z).
\end{equation}
A detailed list of additional, helpful relations for this function can
be found in \cite{44DA}.

The polylogarithm was introduced in \cite{Lou00} to calculate the
thermodynamic properties of the GCM within the MSA where, obviously,
expression (\ref{poly_sum}) was used throughout; this was done even
though there was no guarantee that for certain state points the
modulus of the respective arguments $|z|$ does not exceed 1, violating
thus the condition for the validity of Eq.~(\ref{poly_sum}).  Since
this function plays a central role in the formalism of the MSA and the
SCOZA (see Sec.~\ref{msa} and \ref{scoza}), a reliable evaluation of
${\rm Li}_n(z)$ for arbitrary argument $z$ is indispensable for a
successful solution of the SCOZA-ODE and PDE. We therefore provide in
the following a more detailed presentation of evaluation schemes and
indicate how this function can be calculated in an accurate and
efficient way for arbitrary argument $z$.

In its evaluation of ${\rm Li}_n(z)$, the {\tt MATHEMATICA} software
relies on Euler-MacLaurin summation, expansions in terms of incomplete
Gamma functions, and numerical quadrature \cite{Wol03}. Efficient {\it
and} accurate C- or Fortran-based implementations, on the other hand,
are more difficult to find.  First attempts to evaluate
Eq.~(\ref{poly_n}) directly by various numerical integration schemes
turned out to be either too time-consuming or did not provide results
of sufficient accuracy. Finally, we found that the following
functional relation between the polylogarithm and the complete
Fermi-Dirac function, $F_n(z)$,

\begin{equation}
F_n(z) = \frac{1}{\Gamma(n+1)} \int\limits_0^\infty {\rm d}t \,
\frac{t^n}{{\rm e}^{t-z}+1} = - {\rm Li}_{n+1}(-{\rm e}^z)
\end{equation}
along with the accurate and efficient implementation of $F_n(z)$
via series and asymptotic expansions in combination with Chebyshev
fits, as implemented in the GNU Scientific Library \cite{Galxx}, provided the
desired results, which finally brought the solution of the SCOZA
differential equations within reach.

\section{Acknowledgments}

The authors are indebted to Elisabeth Sch\"oll-Paschinger
(Universit\"at Wien) and Michael Kunzinger (Universit\"at Wien) for
useful discussions and to Mar{\'i}a-Jos{\'e} Fernaud for critical
reading of the manuscript. This work was supported by the
\"Osterreichische Forschungsfonds (FWF) under Project Nos.  P15758 and
P17823. BMM gratefully acknowledges financial support of the Erwin
Schr\"odinger Institute for Mathematical Physics (Wien), where part of
this work was carried out.


\begin{thebibliography}{99}
\bibitem{note2} {Note, that the 'correct' diffusion equation for non-constant diffusivity $D(\varrho,\beta)$ would read
\begin{displaymath}
\frac{\partial u}{\partial \beta} = \frac{\partial}{\partial \varrho}
\left(D(\varrho,\beta) \frac{\partial u}{\partial \varrho}\right).
\end{displaymath}}
\end{thebibliography}
\end{document}